\documentclass[twocolumn,showpacs,preprintnumbers,amsmath,amssymb,prb,superscriptaddress]{revtex4-1}

\usepackage{amsfonts}
\usepackage[english]{babel}
\usepackage[T1]{fontenc}
\usepackage{times}
\usepackage{mathrsfs}
\usepackage{graphicx}% Include figure files
\usepackage{dcolumn}% Align table columns on decimal point
\usepackage{bm}% bold math
\usepackage[colorlinks,bookmarks=false,citecolor=blue,linkcolor=red,urlcolor=blue]{hyperref}

\newcommand{\be}{\begin{eqnarray}}
\newcommand{\ee}{\end{eqnarray}}
\newcommand{\ba}{\begin{align}}
\newcommand{\ea}{\end{align}}

\usepackage{tabularx}

\usepackage[usenames,dvipsnames]{xcolor}
\usepackage{soul}
\setulcolor{BrickRed}
\setstcolor{purple}

\makeatletter
\newcommand*\dashline{\rotatebox[origin=c]{90}{$\dabar@\dabar@\dabar@$}}
\makeatother

\begin{document}

\title{Transport properties of an interacting Majorana chain}

\author{ Zhao Liu}
\affiliation{Dahlem Center for Complex Quantum Systems and Institut fur Theoretische Physik, Freie Universitat Berlin, 14195 Berlin, Germany}
\author{Emil J. Bergholtz}
\affiliation{Department of Physics, Stockholm University, AlbaNova University Center, 106 91 Stockholm, Sweden}
\author{Alessandro Romito}
\affiliation{Department of Physics, Lancaster University, Lancaster LA1 4YB, United Kingdom}
\author{Dganit Meidan}
\affiliation{Department of Physics, Ben-Gurion University of the Negev, Beer-Sheva 84105, Israel}

\date{\today}

\begin{abstract}
We study a one-dimensional (1D) chain of $2N$ Majorana bound states, which interact through a local quartic interaction. This model describes for example the edge physics of a quasi 1D stack of $2N$ Kitaev chains with modified time-reversal symmetry $T\gamma_iT^{-1}=\gamma_i$, which precludes the presence of quadratic coupling. The ground state of our 1D Majorana chain displays a four-fold periodicity in $N$, corresponding to the four distinct topological classes of the stacked Kitaev chains. We analyze the transport properties of the 1D Majorana chain, when probed by local conductors located at its ends. We find that for finite but large $N$, the scattering matrix partially reflects the four-fold periodicity, and the chain exhibits strikingly different transport properties for different chain lengths. In the thermodynamic limit, the 1D Majorana chain hosts a robust many-body zero mode, which indicates that the corresponding stacked two-dimensional bulk system realizes a weak topological phase.
\end{abstract}
\maketitle

\section{Introduction}

Models of interacting Majorana modes provide a simple platform to study novel physical phenomena. Examples range from emergent supersymmetric quantum critical behavior\cite{Rahmani2015} to the physics of black holes \cite{Sachdev1993,Kitaev2015}. One particularly interesting example is the Sachdev-Ye-Kitaev (SYK) model with random all-to-all Majorana interactions \cite{Sachdev1993,Kitaev2015}, which is a calculable model with implications for quantum gravity, quantum information, and quantum chaos \cite{Kitaev2015, Maldacena2016,Jevicki2016,You2016,Banerjee2017,Cotler2017,Davison2017,Jensen2017,Gu2017,Pikulin2017,Witten2016,Gross2017,Affleck2017}.
Recent proposals for the realization of the SYK model would potentially allow to experimentally probe this physics in a solid state setup \cite{Chew2017,Pikulin2017}. 
A variant of the SYK model with short-range strong interactions was suggested to exhibit emergent supersymmetric quantum critical behavior \cite{Rahmani2015}. Interestingly, this short-range model describes excitations on the edge of stacked topological superconducting wires, which is potentially easier to access experimentally.

Transport properties provide a prominent tool to probe topological phases in (quasi) one-dimensional (1D) systems \cite{Alicea2012}. Indeed, conductance measurements have been the first indications of possible topological phases in engineered nanostructures \cite{Mourik2012,Das2012,Nadj-Perge2014,Deng2016}. The presence of gapless modes confined to the system ends affects the scattering of non-interacting fermions when the system is connected to leads. In fact, for non-interacting fermions in a quasi 1D wire, the topological invariant can be formulated in terms of the scattering matrix invariants \cite{Fulga2011}. 

Interactions are known to alter the topological classification of gapped phases as well as their transport properties. By enlarging the phase space, interactions can connect otherwise distinct topological phases and reduce the number of gapped phases \cite{Fidkowski2010,Fidkowski2011a,Turner2011,Gurarie2011,Morimoto2015}. 
One notable example   
occurs in the stack of 1D topological superconductors with a modified time-reversal symmetry $T^2=1$. In that case, the non-interacting system is characterized by a $\mathbb{Z}$ index which counts the number of Majorana modes at its boundary, while interactions reduce the number of gapped phases to eight (labeled by a $\mathbb{Z}_8$ topological index) \cite{Fidkowski2010}.

It is conceptually useful to think of the eight topological subclasses as constructed by stacking $N$ topological superconducting chains forming a slab of finite transverse size $N$ \cite{Kitaev2001}, with $N$ Majorana modes localized at its boundary. Here interactions of finite range can gap the Majorana modes without breaking time-reversal symmetry if $N$ is a multiple of eight, making such systems adiabatically connected to a trivial insulator. However, the remaining nontrivial phases host   a spinless fermion, a Majorana quasi-particle, and a Kramers doublet, respectively,  as boundary excitations. Each of these physical excitations have markedly distinct measurable features which can be detected when coupling the edge of the stacked system to external leads. Remarkably, this allows to formulate the topological index using a scattering matrix approach---even in the presence of interactions\cite{Meidan2014,Meidan2016}.

As long as the number of stacked superconducting chains is kept finite, the $\mathbb{Z}_8$ periodicity of the system, as a topological property, persists independently of the range of interactions. The interaction range introduces a transverse length scale for the stacked quasi 1D system. 
When the interaction range scales with the system size, the boundary is essentially a zero-dimensional dot, resulting for example in the Sachdev-Ye-Kitaev model \cite{Sachdev1993,Kitaev2015}. For a finite-size boundary, the $\mathbb{Z}_8$ periodicity of the system has been shown to emerge in the energy level statistics for random interaction strengths \cite{You2016}.  
In the opposite limit when interactions are short-ranged, the stacked setup is two-dimensional of finite transverse size, with a 1D boundary. In this limit, it is possible to analyze transport properties along the boundary, whose sensitivity to topological phases has proven to be robust even in the presence of interactions. 

In this manuscript we study the transport properties along a 1D Majorana chain with local interactions [Eq.~\eqref{H:majo}]. This provides an effective description of the edge of stacked topological superconducting chains with modified time-reversal symmetry $T^2=1$. We restrict to an even number of Majorana fermions $L=2N$ in the 1D chain, so that the corresponding stacked system consists of fermion excitations localized at a single edge, and falls into four distinct topological classes depending on $N \mod 4$.

We show that the scattering matrix of the 1D Majorana chain partially reflects this four-fold periodicity, even when the chain is macroscopically large.
The different phases of the stacked system are characterized by different scattering properties along the 1D Majorana chain on the boundary, which can be detected in transport measurements. We find that electrons scattered from the leads can be fully reflected, can acquire a $\pi$ phase shift upon reflection, or can be transmitted across the chain, depending on the four-fold periodicity of $N \mod 4$. Though these distinct transport properties do not fully distinguish all different topological phases, their combination with the quantum dimension of the system's state\cite{qdim}, leads to a full classification of the different topological phases.
 What if any of the $\mathbb{Z}_8$ periodic structure persists when the boundary system is macroscopically large?
 Our analysis shows that in the thermodynamic limit our 1D Majorana chain has a two-fold degenerate ground state. 
The presence of this zero mode is  stable for a generic interactions which do not break time-reversal and translational symmetry.
Regarding the 1D Majorana chain as an effective model that emerges on the edge of stacked superconducting chains,
the presence of a robust zero mode at $N\rightarrow \infty $ indicates that the two-dimensional bulk system is a weak interacting topological phase. This is quite remarkable given that its non-interacting analog is topologically trivial. Despite its stability, we show that the  transport properties of the boundary zero mode become progressively harder to detect when the system is coupled to single channel leads. This is due to the fact that the tunneling matrix elements between the two degenerate ground states decay rapidly in the large $N$ limit.

\section{Model and Symmetries}
\begin{figure}
\includegraphics[width=0.45\textwidth]{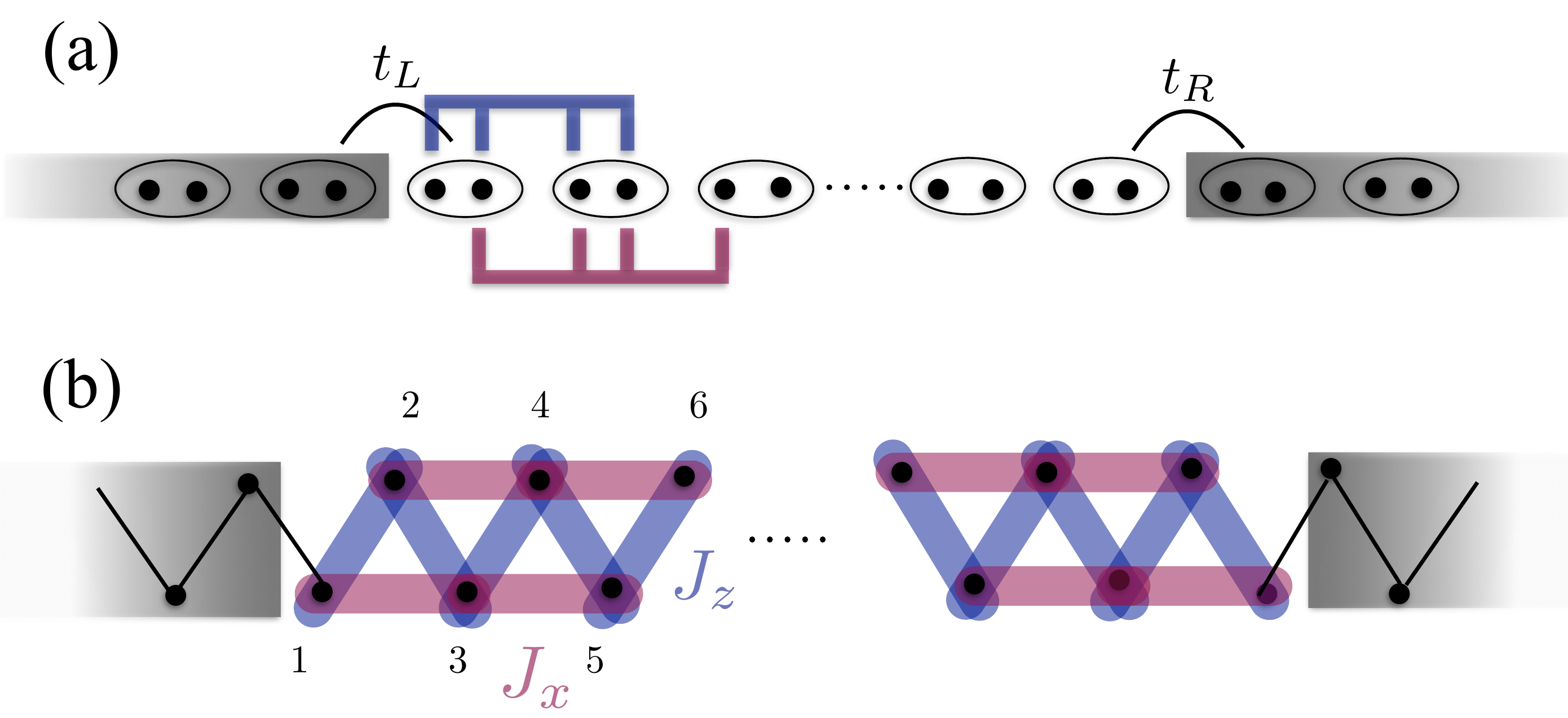}
\caption{(a) Schematics of a 1D chain of Majorana bound states coupled to left and right leads. Circled pairs of Majorana zero modes indicate fermionic degrees of freedom and colored lines represent different possible interaction terms. (b) Equivalent representation of the Majorana chain in (a) after its mapping to a spin model. The different interaction terms in (a) are represented by different spin coupling terms in (b) with the same color code.}
\label{fig:sistema}
\end{figure}

We study a 1D Majorana chain of length $L=2N$ with short-range interactions under open boundary conditions, as described by the Hamiltonian
\be\label{H:majo}
H_0= -W\sum_{i=1}^{2N-3} \gamma_i\gamma_{i+1}\gamma_{i+2}\gamma_{i+3},
\label{eq:hamiltonian}
\ee
where $\gamma_j$ are Majorana bound state operators defined by the algebra $\{\gamma_j,\gamma_k\}=2\delta_{j,k}$ and $\gamma_j=\gamma_j^\dagger$. The system is sketched in Fig.~\ref{fig:sistema} (a).
This 1D chain describes for example the low-energy physics on the edge of a quasi 1D system composed of $2N$ Kitaev chains \cite{Kitaev2001} with modified time-reversal symmetry $T\gamma_iT^{-1}=\gamma_i$  (symmetry class BDI) \cite{Altland1996,Kitaev2009,Schnyder2008}, which precludes the presence of quadratic Majorana terms\cite{Fidkowski2010}. 
 
To assess the symmetry properties of the Hamiltonian \eqref{eq:hamiltonian}, it is convenient to express it in terms of spin degrees of freedom. Following a Jordan-Wigner transformation
\begin{gather}
\gamma_{2n-1} = \prod_{m<n}\sigma_m^z \sigma_n^x ,\nonumber\\
\gamma_{2n} = \prod_{m<n}\sigma^z_m \sigma^y_n ,\nonumber\\
-i\gamma_{2n-1}\gamma_{2n} = \sigma^z_n,\nonumber
\end{gather}
it can be mapped on to a spin chain model
\be\label{H:spin}
H_0 = W\sum_{i=1}^{N-1}\sigma_i^z\sigma_{i+1}^z+W\sum_{i=1}^{N-2} \sigma_i^x\sigma_{i+2}^x,
\ee
which has a natural interpretation in terms of the ladder spin-chain sketched in Fig.~\ref{fig:sistema} (b).

The model has three discrete symmetries.  
The first is a charge conjugation symmetry, $H_0=T_LH_0T_L^{-1}$, which can be represented as:
\be
\nonumber
T_L &=& \left[i\prod_{m=1}^{\lfloor{N/2\rfloor}}\sigma_{2m-1}^{x}\sigma_{2m}^{y}\right]K\\
&=& \left\{
\begin{array}{ccc}
 \left[ \prod_{j=1}^N\gamma_{2j-1}\right]K,  & N\in {\rm odd}      \\
  \left[  \prod_{j=1}^N\gamma_{2j}\right]K,  &N\in {\rm even}
   \end{array}\right.,\nonumber
\label{eq:inversione-temporale}
\ee
where $K$ denotes complex conjugation. Regarding the chain as describing the edge model of a quasi 1D bulk composed of $2N$ Kitaev chains, this operator can be understood as the projection of the global time-reversal symmetry on the low-energy degrees of freedom on the edge \cite{Fidkowski2011a,Turner2011}. Importantly, while the global time-reversal symmetry $T^2=1$, its local projection $T_L^2=\pm1$ depends on the total number of fermionic sites $N$ of the chain.
In addition to charge conjugate, the system possesses two additional symmetries, namely the parity of the odd or the even subchain respectively:
\be
P_{o}&=& \prod_{i}\sigma_{2i-1}^{z} = \prod_i \left(-i\gamma_{4i-3}\gamma_{4i-2}\right),\nonumber\\
P_{e}&=& \prod_{i}\sigma_{2i}^{z} = \prod_i \left(-i\gamma_{4i-1}\gamma_{4i}\right).\nonumber
\ee

\begin{figure}
\centerline{\includegraphics[width=\linewidth]{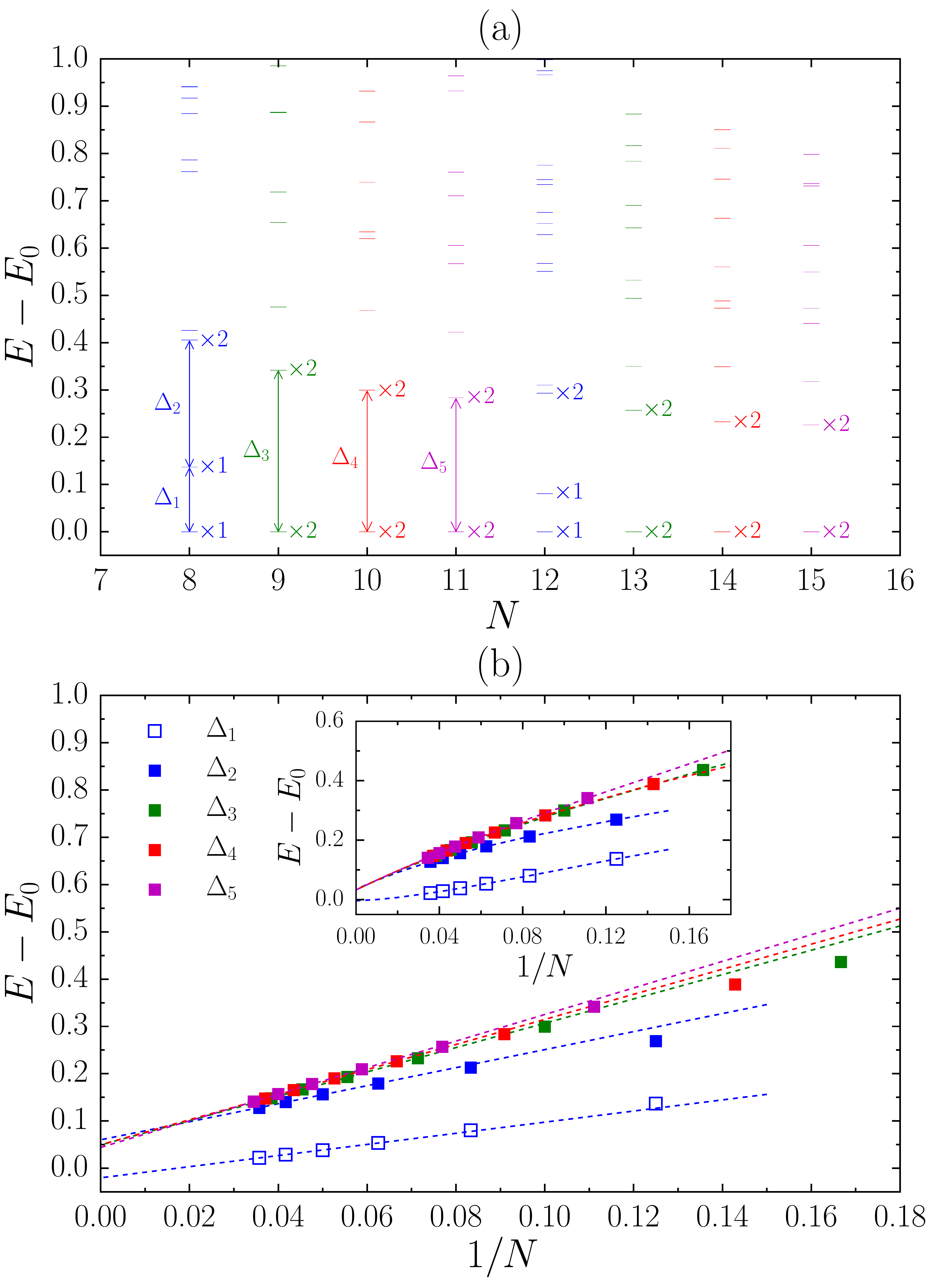}}
\caption{(a) Typical many-body spectra for different system sizes $N=8-15$. The degeneracies of some low-lying levels are indicated by the numbers nearby. For $N\mod 4=1,2,3$, all energy levels are exactly two-fold degenerate. For $N\mod 4=0$, the lowest two levels are non-degenerate, while the third level is again exactly two-fold degenerate. (b) The scaling of the low-energy gaps as defined in (a) with increasing system sizes up to $N=29$. The linear fitting (dashed line) is made according to the data of the largest four system sizes for each $N\mod 4$. The negative $y-$intercept for $\Delta_1$ suggests that the two-fold exact degeneracy of the lowest two levels for $N\mod 4=0$ will be recovered in the thermodynamic limit. The $y-$intercepts for other four gaps have roughly similar nonzero values, suggesting finite gaps in the thermodynamic limit. Intriguingly, as shown in the inset, the unique fifth order polynomial in $1/N$ that goes through all data points for each $N\mod 4$ gives a $y-$intercept for $\Delta_1$ remarkably close to zero while the $\Delta_{2,3,4,5}$'s extrapolate to virtually identical finite values. We set $W=0.5$ in this figure.}
\label{GSDegeneracy}
\end{figure}

Studying the representations of the three discrete symmetries as a function of the number of spins $N$ reveals that the Majorana chain described in Eq.~\eqref{H:spin} falls into four symmetry-protected topological phases labeled by different $\nu = N\mod 4$. Those are distinguished by the sign of $T_L^2$ and by the fact that time-reversal symmetry commutes or anticommutes with the total parity $ P = P_oP_e$ \cite{Fidkowski2011b,Turner2011}. In addition, for the Hamiltonian in Eq. \eqref{H:majo}, the four symmetry-protected phases can be characterized by different subsets of non-commuting symmetry operators: 
\begin{itemize}
\item[(A)] For a chain with  $ \nu=0$, the local time-reversal operator commutes with both even chain and odd chain parities $[T_L,P_e]=[T_L,P_o] = 0 $, and the system has a unique ground state.
\item[(B)] For a chain with $\nu =1 $, $\{T_L, P_o\} =0$, giving rise to a two-fold degeneracy of the ground-state manifold, where the two ground states differ by the parity of the odd chain.
\item[(C)] For a chain with $\nu=2 $,  $\{T_L, P_o\} =\{T_L, P_e\}=0$, and the ground state is two-fold degenerate, where the two ground states differ by the relative parities of the odd and the even chain but have the same total parity. 
\item[(D)] For a chain with $\nu =3 $, $\{T_L, P_e\} =0$, and the ground state is two-fold degenerate, where the two ground states differ by the parity of the even chain.
\end{itemize}
Thus each of the three non-trivial phases ($\nu=1,2,3$) are characterized by a two-fold degenerate ground state, which is spanned by the parity of the even and/or odd subchains (Fig.~\ref{GSDegeneracy}), in contrast to the unique ground state in the trivial phase ($\nu=0$).

\section{Transport and scattering processes}

To study the transport properties of the Majorana chain in Eq.~\eqref{eq:hamiltonian}, we assume that it is coupled to non-interacting leads located at two ends of the system. We consider local tunnelling so that electrons can tunnel from the lead to the end-fermionic site of the chain, as described by the Hamiltonian
$H=H_0+H_T+H_{L}+H_{R}$, where
\be\label{HT}
H_T = \sum_{k} t_L c_{L,k}^\dag d_{L}+t_R c_{R,k}^\dag d_{R} +{\rm h.c.},
\ee 
and 
\be
H_{j=L,R}=\sum_k \epsilon_j(k) c_{j,k}^\dagger c_{j,k}.\nonumber
\ee
Here $ c_{j,k}$ annihilates an electron of energy $\epsilon_j(k)$ with momentum $k $ in lead $ j$, $d_L=(\gamma_1+i\gamma_2)/2$, and $d_R=(\gamma_{2N-1}+ i \gamma_{2N})/2$,  where $L,R$ label the left and right leads [Fig.~\ref{fig:sistema}~(a)]. The current operator at the left lead is 
\begin{equation}
\hat{I}_L = \partial_t \hat{N}_L = -i \sum_{k} t_{L} c_{L,k}^\dag d_{L} +{\rm h.c.}.\nonumber
\end{equation}

In general, the evaluation of the average current $\langle I_{i=L,R} \rangle$ and the low frequency current fluctuations $\mathscr{S}_{ij\in \{L,R\}}=\mathscr{S}_{ij}(\omega=0)=\int dt\,  \langle\left[  I_i(t) - \langle I_i \rangle, I_j(0)- \langle I_j \rangle \right]_+ \rangle$  through the interacting system in Eq.~\eqref{H:majo} encompasses multi-particle as well as energy-non-conserving processes. 
However, at voltage biases and temperature lower than the gap that separates the ground-state manifold from the excited states, i.e., $ eV_L, eV_R \ll \Delta_{\rm g}$, the system is described by a Fermi liquid and the transport properties can be fully characterized by a unitary scattering matrix $\Psi_{\rm out} =\mathcal{S} \Psi_{{\rm in}}$. The scattering states are in the basis  $\Psi_{\rm in}  = (\psi_{L,e}(E),\psi_{L,h}(E),\psi_{R,e}(E),
\psi_{R,h}(E))^T$ of electron ($e$) and hole ($h$) modes at a given energy $\epsilon$ from the Fermi seas of the left and right leads.

At zero temperature the current at the left lead $L$ and the current fluctuations can be expressed as \cite{Blonder1982,Blanter2000,Anantram1996} 
\begin{gather}
I_L= \frac{e}{h} \sum_{a,b=e,h}\sum_{j=L,R} {\rm sgn}(a) \int dE\, \mathcal{A}_{jb;jb}(L,a;E) f_{j,b}(E), \label{current} \\
\begin{split}
&\mathscr{S}_{ij}= \frac{2e^2}{h} \sum_{k,l=L,R}\sum_{a,b,\gamma,\delta=e,h} {\rm sgn}(a) \, {\rm sgn}(b) \\ 
&\int dE \, \mathcal{A}_{k\gamma;l\delta} (i,a;E) \mathcal{A}_{l\delta;k\gamma}(j,b;E) f_{k,\gamma}(E)[1-f_{l,\delta}(E)], \label{noise}
\end{split} 
\end{gather}
where 
$\mathcal{A}_{k\gamma;l\delta}(i, a;E) = \delta_{i,k} \delta_{i,l} \delta_{a,\gamma} \delta_{a,\delta} - {s_{ik}^{a \gamma}}^*(E) s_{il}^{a\delta}(E)$
 with  $s_{ik}^{\delta \gamma} $  the components of the scattering matrix $ {\cal S}$,    $f_{j,b} (E)= \Theta(E- {\rm sgn}(b) eV_j)$ is the zero temperature Fermi-Dirac distribution, and   ${\rm sgn}(a)$ is positive (negative) for $a=e(h)$.
In the following subsections we study the transport properties of the different topological phases. 

\subsection{$\nu =0 $}
A chain with $ N\mod 4 =0 $ realizes a trivial phase with a unique ground state (quantum dimension 1), separated from the excited stated by a finite gap $\Delta_{\rm g}$, as shown in  Fig.~\ref{GSDegeneracy}.  
At low voltage, $eV < \Delta_{\rm g}$, the system resembles   a trivial insulator and the scattering matrix is  $\mathcal{S}(\omega) = \openone_{4\times 4}$. 

\subsection{$ \nu =1 $}

\begin{figure}
\centerline{\includegraphics[width=\linewidth]{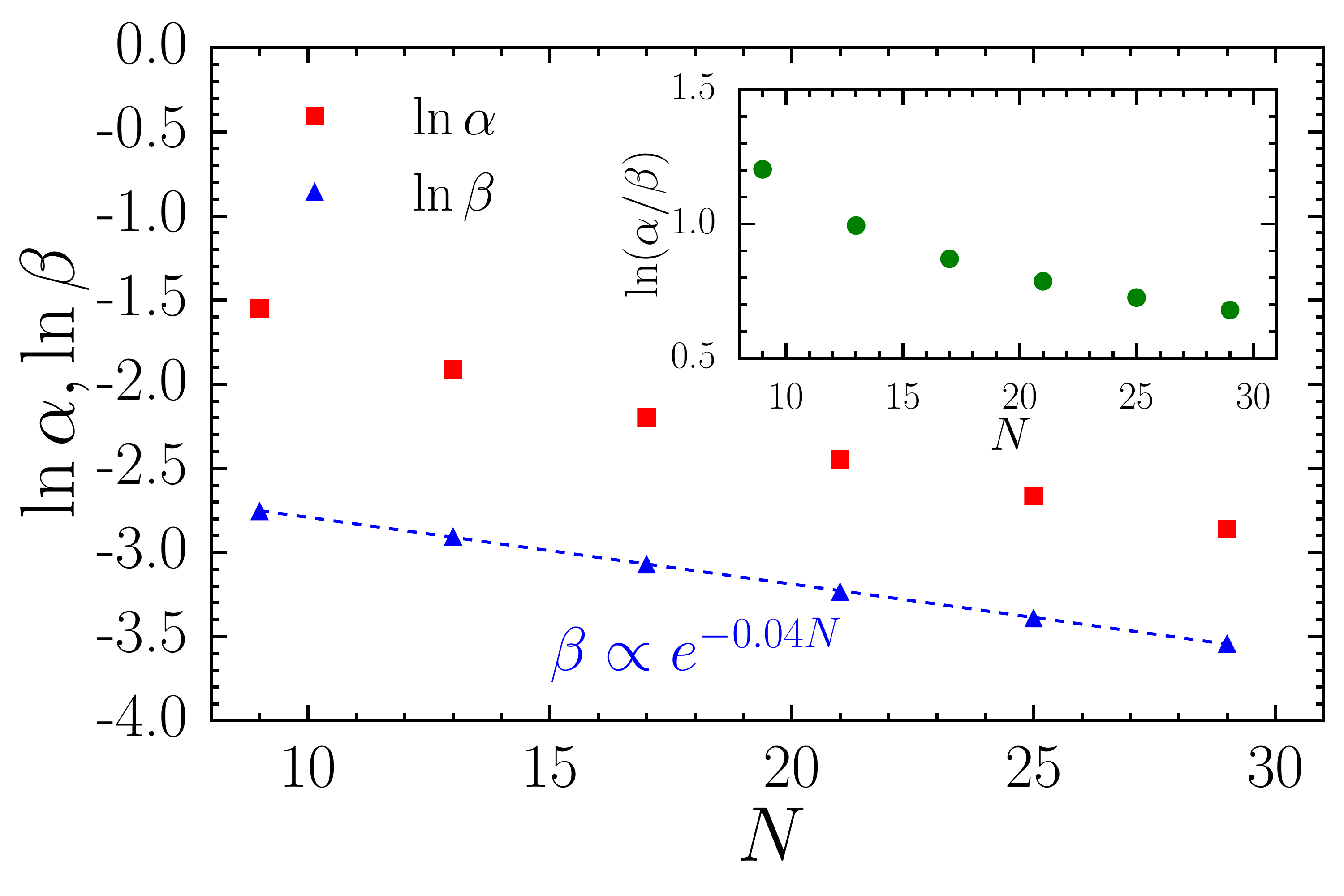}}
\caption{The scaling of $\alpha$ and $\beta$ with increasing system sizes for $N\mod 4=1$ up to $N=29$. The exponential for $\beta\propto e^{-0.04N}$ is obtained  by fitting (dashed line) the data of all system sizes. The precise scaling of $\alpha$ is harder to predict from the available system sizes, although it is most likely exponential in the large $N$ limit. The inset shows the logarithm of the ratio $\alpha/\beta$ which likely approaches a finite value for large $N$, although the existing finite size data do not unambigously confirm this scenario. 
}
\label{scaling_of_tunneling}
\end{figure}

A chain with $N\mod 4=1$ has two degenerate ground states (quantum dimension 2) which are distinguished by the parity of the odd subchain. The ground-state manifold is separated by a finite gap $\Delta_{\rm g}$ from the rest of the spectrum (Fig~\ref{GSDegeneracy}). We distinguish the two ground states by the odd chain parity quantum number  $ P_o|{\rm gs}_{ \pm}\rangle=\pm|{\rm gs}_{ \pm}\rangle $. 
For a chain with $ N\mod 4=1$, tunnelling to and from the two leads changes the parity of the odd subchain and therefore may toggle between the two ground states.  When projecting onto the ground-state manifold, and to the lowest order in the tunneling $t_{L,R} $, the operators $ d_1$ and $d_N $ can be expressed as
\be
{\cal P} d_1{\cal P} &=&\alpha |{\rm gs}_{ +}\rangle \langle {\rm gs}_{-}|+\beta |{\rm gs}_{ -}\rangle \langle {\rm gs}_{+}|\equiv \alpha f^\dag +\beta f, \label{eq:coefficiente1}\\
{\cal P} d_N{\cal P} &=&\tilde{\alpha} |{\rm gs}_{ +}\rangle \langle {\rm gs}_{-}|+\tilde{\beta} |{\rm gs}_{ -}\rangle \langle {\rm gs}_{+}|\equiv \tilde{\alpha} f^\dag +\tilde{\beta} f, \label{eq:coefficiente2}
\ee
where $ {\cal P} = \sum_{s=\pm}|{\rm gs}_{ s}\rangle\langle {\rm gs}_{ s}|$ is the projection operator on the ground-state manifold. Since the system possesses an inversion symmetry $\mathcal{I}$, $\mathcal{I} {\pmb \sigma}_j \mathcal{I}^{-1} ={\pmb \sigma}_{N+1-j}$, which can be written as  $\mathcal{I} d_j \mathcal{I}^{-1} =\left( \prod_{i=1}^{N-j} d_i^\dag d_i\right) d_{N+1-j}$ in terms of the fermions, the coefficients introduced in Eqs.~\eqref{eq:coefficiente1} and \eqref{eq:coefficiente2} satisfy  $ \alpha = \tilde{\alpha}$ and $\beta = -\tilde{\beta} $.  The scaling of matrix elements $ \alpha $ and $ \beta$ with system size is shown in Fig.~\ref{scaling_of_tunneling}. While $\alpha$ and $\beta$ are substantially different in magnitude for any system size leading to interesting consequences in transport, we expect them to decay exponentially with a common exponent as we elaborate on in Section~\ref{thermo}.

When projecting the Hamiltonian on the ground-state manifold, we find
\be
\begin{split}
\tilde{H} & = {\cal P} (H_0+H_T) {\cal P} = {\cal P} H_T {\cal P} \\
 & =  \sum_{k}\sum_{i=L,R} \left(\alpha_i c_{i,k}^\dag f^\dag +\beta_i c_{i,k}^\dag f\right) +{\rm h.c.},
\end{split}\label{eq:proiezione}
\ee
where $ \alpha_{L,R} =  \alpha t_{L,R}$, $ \beta_{L} =\beta  t_{L} $ and $\beta_R=-\beta t_R$.
Eq.~\eqref{eq:proiezione} shows that in the weak tunneling limit, the contributions to the current are dominated by single-particle processes. We can therefore compute the scattering matrix as
\be
\mathcal{S}(\omega)=1-2\pi iW^{\dagger}(\omega-\tilde{H}+i\pi WW^{\dagger})^{-1}W,\nonumber
\ee
where $ W$ is the matrix that describes the coupling of the ground-state manifold of Eq.~\eqref{H:majo} to the leads. 

The resulting scattering matrix generically allows for all possible single-particle scattering channels which are controlled by the coupling to the leads and the interaction-dependent parameters $\alpha$ and $\beta$.
When the leads are coupled symmetrically to the chain $t_L=t_R=t $, the scattering matrix takes a simple form:
\be
{\cal S}(\omega) = \left(
\begin{array}{cccc}
 s^{ee}_{LL} & s^{eh}_{LL}    &s^{ee}_{LR}&0  \\
  s^{he}_{LL}  &      s^{hh}_{LL}&0& s^{hh}_{LR}\\
s^{ee}_{RL}&0&   s^{ee}_{RR} & s^{eh}_{RR}  \\
0 & s^{hh}_{RL}& s^{he}_{RR}  &   s^{hh}_{RR}
\end{array}
\right)\nonumber
\ee
with 
\be
 s^{ee}_{LL} &=& s^{hh}_{LL}= s^{ee}_{RR}= s^{hh}_{RR}= \frac{i \omega }{\Gamma \left(\alpha ^2+\beta ^2\right)+i \omega }, \nonumber\\
 s^{ee}_{LR}&=&s^{ee}_{RL}=\frac{\Gamma  \left(\beta ^2-\alpha ^2\right)}{\Gamma \left(\alpha ^2+\beta ^2\right)+i \omega },\nonumber\\
 s^{he}_{LL }&=& s^{he}_{RR} = s^{eh}_{LL} = s^{eh}_{RR} =\frac{2\Gamma\alpha  \beta }{\Gamma\left(\alpha ^2+\beta ^2\right)+i \omega },\nonumber
  \ee  
  and $ \Gamma = 2\pi t^2$.

From Eq.~\eqref{current}, we obtain the current at the left lead 
\begin{equation}
I_L= \frac{2e^2}{h} \left[ V_L-V_R+\frac{4\alpha^2 \beta^2}{(\alpha^2+\beta^2)^2}(V_L+V_R)\right] .\nonumber
\end{equation}
Similarly, the zero-temperature noise is obtained from Eq.~\eqref{noise} as
\begin{equation}
\mathscr{S}_{LR}(\omega=0)=- \frac{2e^3}{h} \frac{ \left( 2 \alpha \beta (\beta^2-\alpha^2) \right)^2}{\left( \alpha^2+\beta^2 \right)^4} (V_L+V_R).\nonumber
\end{equation}

\subsection{$\nu =2$}

For a chain with  $N\mod 4=2 $, the two degenerate ground states (quantum dimension 2) have opposite fermion parities on both the even and odd subchains. We label the two ground states as $|{\rm gs}_{ +-}\rangle$ and  $|{\rm gs}_{ -+}\rangle$ where the two indices label the parity of the odd and even subchain respectively, and we have assumed that the two ground states have odd total parity $P_e P_o$. Since $H_T$ modifies the parity of only one of the two subchains, it has vanishing matrix elements on the ground-state manifold. Consequently, the low-voltage transport is dominated by virtual transition into the excited states. 

To find how  these higher order processes affect the transport  properties, we perform a Schrieffer Wolff transformation to derive the effective Hamiltonian taking into account virtual transitions to excited states. 
The resulting effective model is  up to an additive constant given by
\begin{widetext}
\be\label{H_SD}
\nonumber
H_{\rm eff} 
&=&\sum_{j=L,R}\sum_{ k}\epsilon_j(k) c^\dag_{j, k} c_{j, k}+\frac{\tau_z}{2}\sum_{k,k'} \left\{\left(J_z+\Delta_z\right)\left [c_{Rk}^\dag c_{Rk'}-\frac{\delta_{k,k'}}{2}\right]-\left(J_z-\Delta_z\right)\left[c_{Lk}^\dag c_{Lk'}-\frac{\delta_{k,k'}}{2}\right]\right\}\nonumber\\
&+&\left\{\tau_+\sum_{k,k'}\left(J_-c_{Rk}^\dag c_{Lk'}+J_+c_{Lk}^\dag c_{Rk'} +\Delta_+ c_{Lk}^\dag c_{Rk'}^\dag+\Delta_-  c_{Rk}c_{Lk'}\right)+{\rm h.c.}\right\}, \label{eq:effettiva}
\ee
\end{widetext}
where we have defined $\tau_z = |{\rm gs}_{ +-}\rangle \langle  {\rm gs}_{ +-}| - |{\rm gs}_{ -+}\rangle \langle  {\rm gs}_{ -+}| $ and $\tau_+=|{\rm gs}_{ +-}\rangle \langle  {\rm gs}_{ -+}|$, and the coefficients are given by:
\be\label{SD_param_1}
J_z &=&  \sum_n\left\{|t_R|^2  \frac{|\tilde{\beta}_n|^2-|\tilde{\alpha}_n|^2}{E_g-E_n}-|t_L|^2  \frac{|\beta_n|^2-|\alpha_n|^2}{E_g-E_n}\right\},\nonumber\\
J_- &=& 2t_Rt_L^*\sum_n\frac{\tilde{\beta}_n\alpha_n^*}{E_g-E_n}\nonumber,\\
J_+ &=&2t_R^*t_L\sum_n \frac{\tilde{\alpha}_n^*\beta_n}{E_g-E_n},\nonumber
\ee
and
\be
\Delta_z&=& \sum_n\left\{|t_R|^2  \frac{|\tilde{\beta}_n|^2-|\tilde{\alpha}_n|^2}{E_g-E_n}+|t_L|^2  \frac{|\beta_n|^2-|\alpha_n|^2}{E_g-E_n}\right\},\nonumber\\
\Delta_+ &=&2t_Rt_L\sum_n\frac{\tilde{\beta}_n\beta_n}{E_g-E_n},\nonumber\\
\Delta_-&=&2t_R^*t_L^*\sum_n\frac{\tilde{\alpha}_n^*\alpha_n^*}{E_g-E_n}. \nonumber
\ee
Here  the sum is over excited states $|n_{s,s'}\rangle$, for which $H_0 |n_{s,s'}\rangle =E_n |n_{s,s'}\rangle$, $P_o  |n_{s,s'}\rangle=s   |n_{s,s'}\rangle$, and $ P_e |n_{s,s'}\rangle=s' |n_{s,s'}\rangle$, 
$E_{g} $ is the ground state and
 \be
\alpha_n &=&\langle {\rm gs}_{ -+}|  d_1|n_{++}\rangle=  \left(\langle {\rm gs}_{ +-}|  d_1^\dag|n_{--}\rangle\right)^*,\nonumber\\
\beta_n &=&\langle {\rm gs}_{ +-}|  d_1|n_{--}\rangle= \left(\langle {\rm gs}_{-+}|  d_1^\dag|n_{++}\rangle\right)^*,\nonumber\\
\tilde{\alpha}_n &=&\langle {\rm gs}_{ -+}|  d_N|n_{--}\rangle= -\left( \langle {\rm gs}_{ +-}|  d_N^\dag|n_{++}\rangle\right)^*,\nonumber\\
\tilde{\beta}_n &=&\langle {\rm gs}_{ +-}|  d_N|n_{++}\rangle=-\left(  \langle {\rm gs}_{-+}|  d_N^\dag|n_{--}\rangle\right)^*,\nonumber
\ee
where the second equality follows from charge conjugation symmetry $T_L $. The symmetry of the system under spatial inversion $ {\cal I}$, which for the case of $\nu=2 $ exchanges the even and odd parities ($ P_e\leftrightarrow P_o$), imposes the constraints:
\be
\alpha_n = \delta_n\tilde{\beta}_n,\nonumber\\
\beta_n =-\delta_n\tilde{\alpha}_n.\nonumber
\ee
where $\delta_n=\pm$  is an $n$-dependent sign.
We evaluate $\alpha_n$ and $\beta_n$ numerically for different system sizes and determine the corresponding values for the coefficients of the effective Hamiltonian, $J_z$, $J_-$, $J_+$, $\Delta_z$, $\Delta_+$, $\Delta_-$, which are reported in Table~\ref{t3}.

\begin{table}
\caption{Values of the coefficients $J_z$, $J_-$, $J_+$, $\Delta_z$, $\Delta_+$, $\Delta_-$ entering the effective Hamiltonian Eq.~\eqref{eq:effettiva} for different $N=4m+2$.
}
\begin{ruledtabular}
\begin{tabular}{lccc}\label{t3}
 &$N=10$&$N=14$&$N=18$\\
\hline
$ J_z / (|t_R|^2+|t_L|^2) $&$1.076385$&$1.071858$&$1.067348$\\
$ J_+/(2t_L t_R^*) $&$0.420549$&$0.333439$&$0.27389$\\
$ J_-/(2t_L^* t_R) $&$0.06542$&$0.0677778$&$0.0657269$\\
$ \Delta_z/(|t_R|^2-|t_L|^2) $&$1.076385$&$1.071858$&$1.067348$\\
$ \Delta_+/(2t_L t_R) $&$0.150967$&$0.141257$&$0.128129$\\
$ \Delta_-/(2t_L^* t_R^*) $&$0.150967$&$0.141257$&$0.128129$
\end{tabular}
\end{ruledtabular}
\end{table}

The model described by Eq.~\eqref{H_SD} is a variant of the compactified two-channel Kondo model whose low-energy physics has been analyzed by the study of RG flow \cite{Coleman1995,Coleman1995a,Meidan2016}. This analysis shows that the low voltage and temperature limit is governed by screening of the ground-state spin degree-of-freedom by the lead electrons, and the physics is that of a one-channel Kondo.
At low temperature the system is described by a Fermi liquid theory and is characterized by a unitary scattering matrix which takes the standard  form: $\mathcal{S} = -\openone_{4\times 4} $\cite{Ng1988}. While this $ \pi$ phase shift does not effect the conductance, it can in principle be detected in phase sensitive interference type of measurement. 

\subsection{$\nu=3$}

A chain with $N\mod 4 =3$ is topological equivalent to a chain with $ N\mod4=-1$, which is the inverse phase of $ N\mod4=1$.  \cite{Meidan2014} 
Here  $\nu $ and $-\nu$  are inverse of each other in the sense that when combined, the phase and its inverse form a trivial phase \cite{comment1}.
The properties of two chains with $N\mod 4 =-1 $ and $N\mod 4 =1 $ strongly resemble each other. 
Both phases are characterized by a two-fold ground-state degeneracy (quantum dimension 2), and a fermionic  zero mode that toggles between them, changing  the total parity of the chain. In addition, a recent study of the energy level statistics in a related model with random long-range interactions  \cite{You2016} showed that the phase $ \nu=1$ and its inverse $ \nu =-1$ have identical energy level statistics. 
Despite these similarities, our model with local interactions presents distinct transport properties for the two phases.

When $N\mod 4 =-1 $, the ground-state manifold is spanned by the parity of the even subchain, while tunnelling events from the left and right leads change the parity of the odd subchain. Therefore, tunneling of electrons to and from the leads does not introduce transitions within the degenerate subspace, in any order in perturbation theory.
The transport properties of the system with $N\mod 4=-1$ sites  follow that of a non-degenerate ground state and the scattering matrix is given by $\mathcal{S} = \openone_{4\times 4} $.

To summarise the discussion above, it demonstrates that, at low voltage bias $ eV\ll \Delta_g$, the interacting Majorana chain of length $ 2N$, is described by a unitary matrix. The form of the scattering matrix together with the ground-state degeneracy (or quantum dimension) allows to fully resolve the four-fold periodicity of the chain.

\section{Thermodynamic limit}\label{thermo}
Regarding  the chain as the boundary  of a stack of topological 1D superconductors, the distinct transport properties described above reflect the different nature of the  zero modes localized at the boundary of the stacked system.
In all of the three non-trivial classes the system  hosts a topologically protected zero mode. This zero mode gives rise to a two-fold degeneracy in the spectrum of the boundary. As this degeneracy is topologically protected, it  must persist even when the boundary system is macroscopically  large. The trivial phase corresponding to $ N\mod 4 = 0$, on the other hand, has a unique non-degenerate ground state for any finite $N$. This rises the question:  which of these paradigms  will reflect the characteristic behavior of the boundary  system in the thermodynamic limit?  A hint to the answer  lies in the observation that while the degeneracy in the three non-trivial phases is protected by topology and {\it cannot} be lifted for any interaction profile, the non-degenerate  ground state in the $ N\mod 4 = 0$ case is a result of a {\it specific} (albeit generic) choice of the interaction Hamiltonian.  (As a counterexample, a chain Hamiltonian with $N\mod 4=0$ and uniform all-to-all interactions is characterized by a two-fold degenerate ground state).

\begin{figure}[h]
\centerline{\includegraphics[width=\linewidth]{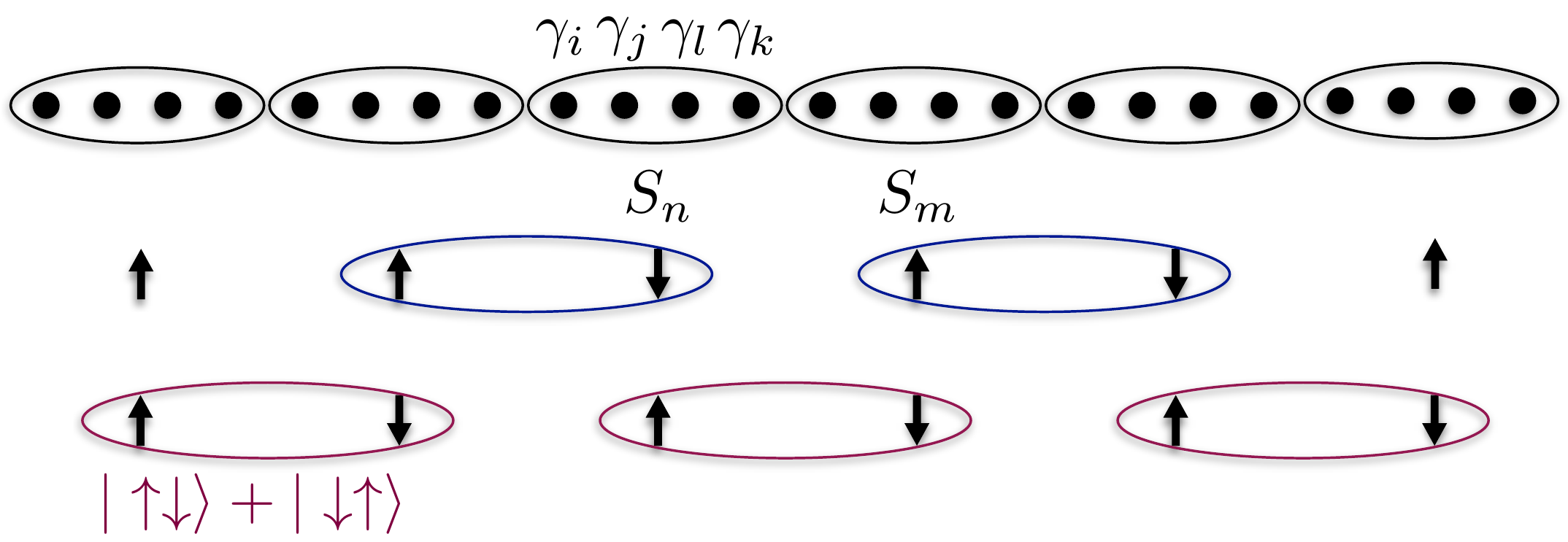}}
\caption{An interacting Majorana chain, divided  into groups of four per site. An intra-site interaction term couples the four Majorana modes leaving a two-fold degeneracy per site corresponding to a spin-$1/2$ chain. Inter-site interaction terms between two neighbouring spins results in a unique spin-singlet ground state. The two ways in which the spin system can be coupled pairwise are topologically distinct. }
\label{spin_chain}
\end{figure}

To address  this question, it is instructive to consider a partition of the Majorana chain into sites consisting of four Majorana modes, as illustrated in Fig.~\ref{spin_chain}. In the absence of interactions, the Majorana operators of each site span a four-fold degenerate ground state. An intra-site interaction term $\gamma_i\gamma_j\gamma_k\gamma_l $ couples the four Majorana modes leaving a two-fold degeneracy per site.  This, in fact, realizes a spin-$1/2$ chain. It can be readily verified that  generic  inter-site interaction terms that couple two neighbouring spins lift the four-fold degeneracy of their respective local Hilbert spaces resulting in a unique ground state of the two-spin system (Fig.~\ref{spin_chain}). The two ways in which the local spins can be dimerized in pairs are topologically distinct, and the interface between them hosts a local spin-$1/2$ zero mode.  We conclude  by noting that the constant interaction profile chosen in our model  lies at the phase boundary between these two distinct phases. It is therefore characterized by the presence of an extended (gapless) mode along the 1D chain.  The energy of the extended mode scales inversely with the system size. In the thermodynamic limit, the excitation energy of this extended mode goes to zero and the ground state becomes doubly degenerate.  This  is analogous to the emergence of a weak topological phase in a non-interacting model of stacked topological insulators \cite{Ringel2012}.  This picture is supported by the scaling of the first excited state energy with system size (Fig.~\ref{GSDegeneracy}).

\begin{figure}
\centerline{\includegraphics[width=\linewidth]{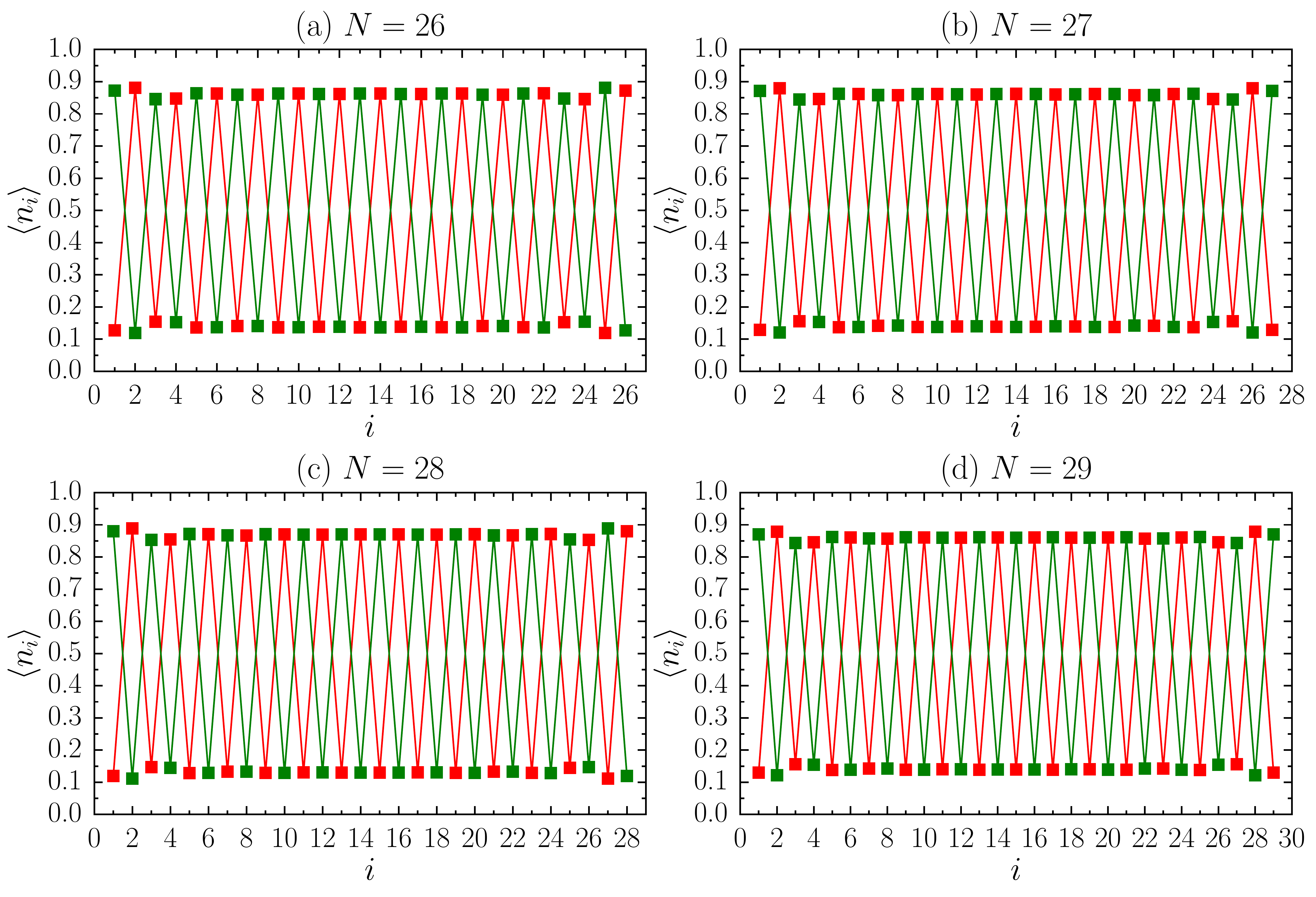}}
\caption{The local density profile  $\langle n_i\rangle= \langle \Psi _{1(2)} |d_i^\dag d_i | \Psi _{1(2)}\rangle $ shown in green (red), respectively,  for different system sizes ranging from $N=26 $ to $N=29$. For $N=26,27$ and $29$, $| \Psi _{1(2)}\rangle $ is the exactly two-fold degenerate ground state. For $N=28$, we use $| \Psi _{1(2)}\rangle=(| \tilde{\Psi} _{1}\rangle\pm| \tilde{\Psi} _{2}\rangle)/\sqrt{2}$, where $|\tilde{\Psi}_{1(2)}\rangle$ are the lowest two eigenstates of the Hamiltonian \eqref{eq:hamiltonian} separated by an energy splitting for finite-size chains. For each $N$, $| \Psi _{1(2)}\rangle $ form shifted charge density waves. 
}
\label{CDW}
\end{figure}

While the boundary zero mode remains stable for a generic interaction which does not break time-reversal and translational symmetry,  its transport properties become progressively  undetectable when the system is coupled to single channel leads.  This is because the tunneling matrix element between the two degenerate ground states vanishes exponentially in this limit (Fig.~\ref{scaling_of_tunneling}). The reason for this exponential suppression can be seen in Fig. \ref{CDW}, which shows that  the density profile of the two ground states form shifted charge density waves.  Such a density profile  indicates that the two ground states are expected to have exponentially vanishing matrix elements upon flipping the occupation locally at the system's end. 

\section{Conclusions}

We have studied the transport properties of an interacting Majorana chain of length $ 2N$ with local interactions and coupled to external metallic leads. The model describes for example the low-energy excitations on the edge of stacked topological superconducting chains with a modified time-reversal symmetry. This stacked system falls into four topological classes depending on $N\mod 4 $. We show that at low voltage bias, the transport properties of the 1D Majorana chain on the edge are characterized by a unitary scattering matrix, which, for finite but large $N$, partially reflects this four-fold periodicity. As a consequence the chain exhibits strikingly different transport properties for different $N\mod 4$. The four-periodicity of the ground states can be fully identified by combining the transport properties with the quantum dimension of the system.

We further provide evidence that in the thermodynamic limit the chain has a two-fold degenerate ground state. Regarding the Majorana chain as an effective model that emerges at the end of a system of stacked superconducting chains, the emergence of a robust mode indicates that  the bulk two-dimensional system is in a weak interacting topological phase. Our finite-size numerics suggest that the two degenerate ground states form shifted charge density waves, indicating that the coupling to the leads vanishes exponentially with increasing system size. 

\section{Acknowledgments}
We acknowledge fruitful discussions in earlier stages of the work with Ehud Altman,  Zohar Nussinov, Jonathan Ruhman, and Felix von Oppen.
Z.L. was supported by Alexander von Humboldt Research Fellowship for Postdoctoral Researchers and the US Department of Energy, Office of Basic Energy Sciences through Grant No. DE-SC0002140. The latter was specifically for the use of computational facilities at Princeton
University. E.J.B was supported by the Wallenberg Academy Fellows program of the Knut and Alice Wallenberg Foundation. A.R. acknowledges support by EPSRC via Grant No. EP/P010180/1. D. M. acknowledges  support from the  Israel Science Foundation (Grant No.~737/14) and
from the People Programme (Marie Curie Actions) of the European Union's Seventh Framework Programme (FP7/2007-2013) under REA grant agreement No. 631064.

%\bibliographystyle{nourl_apsrev}
%\bibliography{tqi-Majorana-interactions}

\begin{thebibliography}{40}
\expandafter\ifx\csname natexlab\endcsname\relax\def\natexlab#1{#1}\fi
\expandafter\ifx\csname bibnamefont\endcsname\relax
  \def\bibnamefont#1{#1}\fi
\expandafter\ifx\csname bibfnamefont\endcsname\relax
  \def\bibfnamefont#1{#1}\fi
\expandafter\ifx\csname citenamefont\endcsname\relax
  \def\citenamefont#1{#1}\fi
\providecommand{\bibinfo}[2]{#2}

\bibitem[{\citenamefont{Rahmani et~al.}(2015)\citenamefont{Rahmani, Zhu, Franz,
  and Affleck}}]{Rahmani2015}
\bibinfo{author}{\bibfnamefont{A.}~\bibnamefont{Rahmani}},
  \bibinfo{author}{\bibfnamefont{X.}~\bibnamefont{Zhu}},
  \bibinfo{author}{\bibfnamefont{M.}~\bibnamefont{Franz}}, \bibnamefont{and}
  \bibinfo{author}{\bibfnamefont{I.}~\bibnamefont{Affleck}}, 
%pp. \bibinfo{pages}{1--14} 
 \eprint{arXiv:1505.03966},
(\bibinfo{year}{2015}).

\bibitem[{\citenamefont{Sachdev and Ye}(1993)}]{Sachdev1993}
\bibinfo{author}{\bibfnamefont{S.}~\bibnamefont{Sachdev}} \bibnamefont{and}
  \bibinfo{author}{\bibfnamefont{J.}~\bibnamefont{Ye}}, \bibinfo{journal}{Phys.
  Rev. Lett.} \textbf{\bibinfo{volume}{70}}, \bibinfo{pages}{3339}
  (\bibinfo{year}{1993})
%, \eprint{9212030}
.

\bibitem{Kitaev2015}
A. Kitaev, http://online.kitp.ucsb.edu/online/entangled15/
kitaev/, http://online.kitp.ucsb.edu/online/entangled15/
kitaev2/ (2015).
%\bibitem[{\citenamefont{Kitaev}(2015)}]{Kitaev2015}
%\bibinfo{author}{\bibfnamefont{A.}~\bibnamefont{Kitaev}},
%  \emph{\bibinfo{title}{{talk at KITP Program: Entanglement in
%  Strongly-Correlated Quantum Matter}}} (\bibinfo{year}{2015}).

\bibitem[{\citenamefont{Maldacena and Stanford}(2016)}]{Maldacena2016}
\bibinfo{author}{\bibfnamefont{J.}~\bibnamefont{Maldacena}} \bibnamefont{and}
  \bibinfo{author}{\bibfnamefont{D.}~\bibnamefont{Stanford}},
  \bibinfo{journal}{Phys. Rev. D} \textbf{\bibinfo{volume}{94}},
  \bibinfo{pages}{106002} (\bibinfo{year}{2016}).
%, \eprint{1604.07818}


\bibitem[{\citenamefont{Jevicki and Suzuki}(2016)}]{Jevicki2016}
\bibinfo{author}{\bibfnamefont{A.}~\bibnamefont{Jevicki}} \bibnamefont{and}
  \bibinfo{author}{\bibfnamefont{K.}~\bibnamefont{Suzuki}},
  \bibinfo{journal}{J. High Energy Phys.} \textbf{\bibinfo{volume}{2016}},
  \bibinfo{pages}{46} (\bibinfo{year}{2016}).
%, \eprint{1608.07567}


\bibitem[{\citenamefont{You et~al.}(2017)\citenamefont{You, Ludwig, and
  Xu}}]{You2016}
\bibinfo{author}{\bibfnamefont{Y.-Z.} \bibnamefont{You}},
  \bibinfo{author}{\bibfnamefont{A.~W.~W.} \bibnamefont{Ludwig}},
  \bibnamefont{and} \bibinfo{author}{\bibfnamefont{C.}~\bibnamefont{Xu}},
  \bibinfo{journal}{Phys. Rev. B} \textbf{\bibinfo{volume}{95}},
  \bibinfo{pages}{115150} (\bibinfo{year}{2017}).
%, \eprint{1602.06964}


\bibitem[{\citenamefont{Banerjee and Altman}(2017)}]{Banerjee2017}
\bibinfo{author}{\bibfnamefont{S.}~\bibnamefont{Banerjee}} \bibnamefont{and}
  \bibinfo{author}{\bibfnamefont{E.}~\bibnamefont{Altman}},
  \bibinfo{journal}{Phys. Rev. B} \textbf{\bibinfo{volume}{95}},
  \bibinfo{pages}{134302} (\bibinfo{year}{2017}).
%, \eprint{1610.04619}


\bibitem[{\citenamefont{Cotler et~al.}(2017)\citenamefont{Cotler, Gur-Ari,
  Hanada, Polchinski, Saad, Shenker, Stanford, Streicher, and
  Tezuka}}]{Cotler2017}
\bibinfo{author}{\bibfnamefont{J.~S.} \bibnamefont{Cotler}},
  \bibinfo{author}{\bibfnamefont{G.}~\bibnamefont{Gur-Ari}},
  \bibinfo{author}{\bibfnamefont{M.}~\bibnamefont{Hanada}},
  \bibinfo{author}{\bibfnamefont{J.}~\bibnamefont{Polchinski}},
  \bibinfo{author}{\bibfnamefont{P.}~\bibnamefont{Saad}},
  \bibinfo{author}{\bibfnamefont{S.~H.} \bibnamefont{Shenker}},
  \bibinfo{author}{\bibfnamefont{D.}~\bibnamefont{Stanford}},
  \bibinfo{author}{\bibfnamefont{A.}~\bibnamefont{Streicher}},
  \bibnamefont{and} \bibinfo{author}{\bibfnamefont{M.}~\bibnamefont{Tezuka}},
  \bibinfo{journal}{J. High Energy Phys.} \textbf{\bibinfo{volume}{2017}},
  \bibinfo{pages}{118} (\bibinfo{year}{2017}).
%, \eprint{1611.04650}


\bibitem[{\citenamefont{Davison et~al.}(2017)\citenamefont{Davison, Fu,
  Georges, Gu, Jensen, and Sachdev}}]{Davison2017}
\bibinfo{author}{\bibfnamefont{R.~A.} \bibnamefont{Davison}},
  \bibinfo{author}{\bibfnamefont{W.}~\bibnamefont{Fu}},
  \bibinfo{author}{\bibfnamefont{A.}~\bibnamefont{Georges}},
  \bibinfo{author}{\bibfnamefont{Y.}~\bibnamefont{Gu}},
  \bibinfo{author}{\bibfnamefont{K.}~\bibnamefont{Jensen}}, \bibnamefont{and}
  \bibinfo{author}{\bibfnamefont{S.}~\bibnamefont{Sachdev}},
  \bibinfo{journal}{Phys. Rev. B} \textbf{\bibinfo{volume}{95}},
  \bibinfo{pages}{155131} (\bibinfo{year}{2017}).
%, \eprint{1612.00849}


\bibitem[{\citenamefont{Jensen}(2016)}]{Jensen2017}
\bibinfo{author}{\bibfnamefont{K.}~\bibnamefont{Jensen}},
  \bibinfo{journal}{Phys. Rev. Lett.} \textbf{\bibinfo{volume}{117}},
  \bibinfo{pages}{111601} (\bibinfo{year}{2016}).
%, \eprint{1605.06098}


\bibitem[{\citenamefont{Gu et~al.}(2017)\citenamefont{Gu, Qi, and
  Stanford}}]{Gu2017}
\bibinfo{author}{\bibfnamefont{Y.}~\bibnamefont{Gu}},
  \bibinfo{author}{\bibfnamefont{X.-L.} \bibnamefont{Qi}}, \bibnamefont{and}
  \bibinfo{author}{\bibfnamefont{D.}~\bibnamefont{Stanford}},
  \bibinfo{journal}{J. High Energy Phys.} \textbf{\bibinfo{volume}{2017}},
  \bibinfo{pages}{125} (\bibinfo{year}{2017}).
%, \eprint{1609.07832}


\bibitem[{\citenamefont{Pikulin and Franz}(2017)}]{Pikulin2017}
\bibinfo{author}{\bibfnamefont{D.~I.} \bibnamefont{Pikulin}} \bibnamefont{and}
  \bibinfo{author}{\bibfnamefont{M.}~\bibnamefont{Franz}}, 
%pp. \bibinfo{pages}{1--16} , 
\eprint{arXiv:1702.04426},
(\bibinfo{year}{2017}).

\bibitem[{\citenamefont{Witten}(2016)}]{Witten2016}
\bibinfo{author}{\bibfnamefont{E.}~\bibnamefont{Witten}}, 
\eprint{arXiv:1610.09758} (\bibinfo{year}{2016}).

\bibitem[{\citenamefont{Gross and Rosenhaus}(2017)}]{Gross2017}
\bibinfo{author}{\bibfnamefont{D.~J.} \bibnamefont{Gross}} \bibnamefont{and}
  \bibinfo{author}{\bibfnamefont{V.}~\bibnamefont{Rosenhaus}},
  \bibinfo{journal}{J. High Energy Phys.} \textbf{\bibinfo{volume}{2017}},
  \bibinfo{pages}{93} (\bibinfo{year}{2017}).
%, \eprint{1610.01569}

  \bibitem{Affleck2017} I. Affleck, A. Rahmani, and D. Pikulin, arXiv:1706.05469 (2017).

\bibitem[{\citenamefont{Chew et~al.}(2017)\citenamefont{Chew, Essin, and
  Alicea}}]{Chew2017}
\bibinfo{author}{\bibfnamefont{A.}~\bibnamefont{Chew}},
  \bibinfo{author}{\bibfnamefont{A.}~\bibnamefont{Essin}}, \bibnamefont{and}
  \bibinfo{author}{\bibfnamefont{J.}~\bibnamefont{Alicea}},
% pp. \bibinfo{pages}{1--7} , 
\eprint{arXiv:1703.06890} (\bibinfo{year}{2017}).

\bibitem[{\citenamefont{Alicea}(2012)}]{Alicea2012}
\bibinfo{author}{\bibfnamefont{J.}~\bibnamefont{Alicea}},
  \bibinfo{journal}{Reports Prog. Phys.} \textbf{\bibinfo{volume}{75}},
  \bibinfo{pages}{76501} (\bibinfo{year}{2012}).
%, \eprint{1202.1293}


\bibitem[{\citenamefont{Mourik et~al.}(2012)\citenamefont{Mourik, Zuo, Frolov,
  Plissard, Bakkers, and Kouwenhoven}}]{Mourik2012}
\bibinfo{author}{\bibfnamefont{V.}~\bibnamefont{Mourik}},
  \bibinfo{author}{\bibfnamefont{K.}~\bibnamefont{Zuo}},
  \bibinfo{author}{\bibfnamefont{S.~M.} \bibnamefont{Frolov}},
  \bibinfo{author}{\bibfnamefont{S.~R.} \bibnamefont{Plissard}},
  \bibinfo{author}{\bibfnamefont{E.~P. a.~M.} \bibnamefont{Bakkers}},
  \bibnamefont{and} \bibinfo{author}{\bibfnamefont{L.~P.}
  \bibnamefont{Kouwenhoven}}, \bibinfo{journal}{Science}
  \textbf{\bibinfo{volume}{336}}, \bibinfo{pages}{1003} (\bibinfo{year}{2012}).
%, \eprint{1204.2792}


\bibitem[{\citenamefont{Das et~al.}(2012)\citenamefont{Das, Ronen, Most, Oreg,
  Heiblum, and Shtrikman}}]{Das2012}
\bibinfo{author}{\bibfnamefont{A.}~\bibnamefont{Das}},
  \bibinfo{author}{\bibfnamefont{Y.}~\bibnamefont{Ronen}},
  \bibinfo{author}{\bibfnamefont{Y.}~\bibnamefont{Most}},
  \bibinfo{author}{\bibfnamefont{Y.}~\bibnamefont{Oreg}},
  \bibinfo{author}{\bibfnamefont{M.}~\bibnamefont{Heiblum}}, \bibnamefont{and}
  \bibinfo{author}{\bibfnamefont{H.}~\bibnamefont{Shtrikman}},
  \bibinfo{journal}{Nat. Phys.} \textbf{\bibinfo{volume}{8}},
  \bibinfo{pages}{887} (\bibinfo{year}{2012}).
%, \eprint{1205.7073}


\bibitem[{\citenamefont{Nadj-Perge et~al.}(2014)\citenamefont{Nadj-Perge,
  Drozdov, Li, Chen, Jeon, Seo, MacDonald, Bernevig, and
  Yazdani}}]{Nadj-Perge2014}
\bibinfo{author}{\bibfnamefont{S.}~\bibnamefont{Nadj-Perge}},
  \bibinfo{author}{\bibfnamefont{I.~K.} \bibnamefont{Drozdov}},
  \bibinfo{author}{\bibfnamefont{J.}~\bibnamefont{Li}},
  \bibinfo{author}{\bibfnamefont{H.}~\bibnamefont{Chen}},
  \bibinfo{author}{\bibfnamefont{S.}~\bibnamefont{Jeon}},
  \bibinfo{author}{\bibfnamefont{J.}~\bibnamefont{Seo}},
  \bibinfo{author}{\bibfnamefont{A.~H.} \bibnamefont{MacDonald}},
  \bibinfo{author}{\bibfnamefont{B.~A.} \bibnamefont{Bernevig}},
  \bibnamefont{and} \bibinfo{author}{\bibfnamefont{A.}~\bibnamefont{Yazdani}},
  \bibinfo{journal}{Science} \textbf{\bibinfo{volume}{346}},
  \bibinfo{pages}{602} (\bibinfo{year}{2014}).
%, \eprint{1410.0682}


\bibitem[{\citenamefont{Deng et~al.}(2016)\citenamefont{Deng, S. Vaitiek\.{e}nas,
  Hansen, Danon, Leijnse, Flensberg, Nyg\r{a}rd, Krogstrup, and Marcus}}]{Deng2016}
\bibinfo{author}{\bibfnamefont{M.~T.} \bibnamefont{Deng}},
  \bibinfo{author}{\bibfnamefont{S.}~\bibnamefont{Vaitiek\.{e}nas}},
  \bibinfo{author}{\bibfnamefont{E.~B.} \bibnamefont{Hansen}},
  \bibinfo{author}{\bibfnamefont{J.}~\bibnamefont{Danon}},
  \bibinfo{author}{\bibfnamefont{M.}~\bibnamefont{Leijnse}},
  \bibinfo{author}{\bibfnamefont{K.}~\bibnamefont{Flensberg}},
  \bibinfo{author}{\bibfnamefont{J.}~\bibnamefont{Nyg\r{a}rd}},
  \bibinfo{author}{\bibfnamefont{P.}~\bibnamefont{Krogstrup}},
  \bibnamefont{and} \bibinfo{author}{\bibfnamefont{C.~M.}
  \bibnamefont{Marcus}}, \bibinfo{journal}{Science}
  \textbf{\bibinfo{volume}{354}}, \bibinfo{pages}{1557} (\bibinfo{year}{2016}).
%, \eprint{1612.07989}


\bibitem[{\citenamefont{Fulga et~al.}(2011)\citenamefont{Fulga, Hassler,
  Akhmerov, and Beenakker}}]{Fulga2011}
\bibinfo{author}{\bibfnamefont{I.~C.} \bibnamefont{Fulga}},
  \bibinfo{author}{\bibfnamefont{F.}~\bibnamefont{Hassler}},
  \bibinfo{author}{\bibfnamefont{A.~R.} \bibnamefont{Akhmerov}},
  \bibnamefont{and} \bibinfo{author}{\bibfnamefont{C.~W.~J.}
  \bibnamefont{Beenakker}}, \bibinfo{journal}{Phys. Rev. B}
  \textbf{\bibinfo{volume}{83}}, \bibinfo{pages}{155429}
  (\bibinfo{year}{2011}).
%, \eprint{1101.1749}


\bibitem[{\citenamefont{Fidkowski and Kitaev}(2010)}]{Fidkowski2010}
\bibinfo{author}{\bibfnamefont{L.}~\bibnamefont{Fidkowski}} \bibnamefont{and}
  \bibinfo{author}{\bibfnamefont{A.}~\bibnamefont{Kitaev}},
  \bibinfo{journal}{Phys. Rev. B} \textbf{\bibinfo{volume}{81}},
  \bibinfo{pages}{134509} (\bibinfo{year}{2010}).
%, \eprint{0904.2197}


\bibitem[{\citenamefont{Fidkowski and Kitaev}(2011)}]{Fidkowski2011a}
\bibinfo{author}{\bibfnamefont{L.}~\bibnamefont{Fidkowski}} \bibnamefont{and}
  \bibinfo{author}{\bibfnamefont{A.}~\bibnamefont{Kitaev}},
  \bibinfo{journal}{Phys. Rev. B} \textbf{\bibinfo{volume}{83}},
  \bibinfo{pages}{75103} (\bibinfo{year}{2011}).
%, \eprint{1008.4138}


\bibitem[{\citenamefont{Turner et~al.}(2011)\citenamefont{Turner, Pollmann, and
  Berg}}]{Turner2011}
\bibinfo{author}{\bibfnamefont{A.~M.} \bibnamefont{Turner}},
  \bibinfo{author}{\bibfnamefont{F.}~\bibnamefont{Pollmann}}, \bibnamefont{and}
  \bibinfo{author}{\bibfnamefont{E.}~\bibnamefont{Berg}},
  \bibinfo{journal}{Phys. Rev. B} \textbf{\bibinfo{volume}{83}},
  \bibinfo{pages}{75102} (\bibinfo{year}{2011}).
%, \eprint{1008.4346}


\bibitem[{\citenamefont{Gurarie}(2011)}]{Gurarie2011}
\bibinfo{author}{\bibfnamefont{V.}~\bibnamefont{Gurarie}},
  \bibinfo{journal}{Phys. Rev. B} \textbf{\bibinfo{volume}{83}},
  \bibinfo{pages}{85426} (\bibinfo{year}{2011}).
%, \eprint{1011.2273}


\bibitem[{\citenamefont{Morimoto et~al.}(2015)\citenamefont{Morimoto, Furusaki,
  and Mudry}}]{Morimoto2015}
\bibinfo{author}{\bibfnamefont{T.}~\bibnamefont{Morimoto}},
  \bibinfo{author}{\bibfnamefont{A.}~\bibnamefont{Furusaki}}, \bibnamefont{and}
  \bibinfo{author}{\bibfnamefont{C.}~\bibnamefont{Mudry}},
  \bibinfo{journal}{Phys. Rev. B} \textbf{\bibinfo{volume}{92}},
  \bibinfo{pages}{125104} (\bibinfo{year}{2015}).
%, \eprint{1505.06341}


\bibitem[{\citenamefont{Kitaev}(2001)}]{Kitaev2001}
\bibinfo{author}{\bibfnamefont{A.}~\bibnamefont{Kitaev}},
  \bibinfo{journal}{Physics-Uspekhi} \textbf{\bibinfo{volume}{44}},
  \bibinfo{pages}{16} (\bibinfo{year}{2001}).
%, \eprint{0010440}


\bibitem[{\citenamefont{Meidan et~al.}(2014)\citenamefont{Meidan, Romito, and
  Brouwer}}]{Meidan2014}
\bibinfo{author}{\bibfnamefont{D.}~\bibnamefont{Meidan}},
  \bibinfo{author}{\bibfnamefont{A.}~\bibnamefont{Romito}}, \bibnamefont{and}
  \bibinfo{author}{\bibfnamefont{P.~W.} \bibnamefont{Brouwer}},
  \bibinfo{journal}{Phys. Rev. Lett.} \textbf{\bibinfo{volume}{113}},
  \bibinfo{pages}{57003} (\bibinfo{year}{2014}).
%, \eprint{arXiv:1312.6367v1}

\bibitem{comment1}
The negative values of $\nu$ can be understood as follows. For a non-interacting chain of Majorana fermions with time-reversal symmetry $T^2=1$, the Majorana zero modes can be classified as even or odd under the action of time reversal symmetry according to $T \gamma_j T^{-1}=\pm  \gamma_j$. In a non-interacting chain hosting both even and odd Majorana zero modes, quadratic coupling terms pairing Majoranas with different parity are allowed, and the topological index $\nu \in \mathbb{Z}= N_e-N_o$ is given by the number of unpaired Majorana zero modes. Here $\nu<0$ stands for $N_o =\nu$ odd Majorana modes. In this representation, the eight distinct topological phases are given by $\nu \in [-3,4]$. For chains with $2N$ Majorana modes considered here, this yields $N \mod 4  \in [-1,2]$, and the $\nu=3$ case is consistently identified with the $\nu=-1$. 


\bibitem[{\citenamefont{Meidan et~al.}(2016)\citenamefont{Meidan, Romito, and
  Brouwer}}]{Meidan2016}
\bibinfo{author}{\bibfnamefont{D.}~\bibnamefont{Meidan}},
  \bibinfo{author}{\bibfnamefont{A.}~\bibnamefont{Romito}}, \bibnamefont{and}
  \bibinfo{author}{\bibfnamefont{P.~W.} \bibnamefont{Brouwer}},
  \bibinfo{journal}{Phys. Rev. B} \textbf{\bibinfo{volume}{93}},
  \bibinfo{pages}{125433} (\bibinfo{year}{2016}).
%, \eprint{1512.04278}


\bibitem{qdim} In this context the quantum dimension is defined as the ground-state degeneracy.


\bibitem[{\citenamefont{Altland and Zirnbauer}(1997)}]{Altland1996}
\bibinfo{author}{\bibfnamefont{A.}~\bibnamefont{Altland}} \bibnamefont{and}
  \bibinfo{author}{\bibfnamefont{M.~R.} \bibnamefont{Zirnbauer}},
  \bibinfo{journal}{Phys. Rev. B} \textbf{\bibinfo{volume}{55}},
  \bibinfo{pages}{1142} (\bibinfo{year}{1997}).
%, \eprint{9602137}


\bibitem{Kitaev2009}
 A. Kitaev, V. Lebedev, and  M. Feigelman,
%Periodic table for topological insulators and superconductors. 
in AIP Conference Proceedings, {\bf 22},  22. (2009).

\bibitem[{\citenamefont{Schnyder et~al.}(2008)\citenamefont{Schnyder, Ryu,
  Furusaki, and Ludwig}}]{Schnyder2008}
\bibinfo{author}{\bibfnamefont{A.~P.} \bibnamefont{Schnyder}},
  \bibinfo{author}{\bibfnamefont{S.}~\bibnamefont{Ryu}},
  \bibinfo{author}{\bibfnamefont{A.}~\bibnamefont{Furusaki}}, \bibnamefont{and}
  \bibinfo{author}{\bibfnamefont{A.~W.~W.} \bibnamefont{Ludwig}},
  \bibinfo{journal}{Phys. Rev. B} \textbf{\bibinfo{volume}{78}},
  \bibinfo{pages}{195125} (\bibinfo{year}{2008}).
%, \eprint{0905.2029}


\bibitem[{\citenamefont{Fidkowski et~al.}(2011)\citenamefont{Fidkowski,
  Lutchyn, Nayak, and Fisher}}]{Fidkowski2011b}
\bibinfo{author}{\bibfnamefont{L.}~\bibnamefont{Fidkowski}},
  \bibinfo{author}{\bibfnamefont{R.~M.} \bibnamefont{Lutchyn}},
  \bibinfo{author}{\bibfnamefont{C.}~\bibnamefont{Nayak}}, \bibnamefont{and}
  \bibinfo{author}{\bibfnamefont{M.~P.~A.} \bibnamefont{Fisher}},
  \bibinfo{journal}{Phys. Rev. B} \textbf{\bibinfo{volume}{84}},
  \bibinfo{pages}{195436} (\bibinfo{year}{2011}).
%, \eprint{1106.2598}


\bibitem[{\citenamefont{Blonder et~al.}(1982)\citenamefont{Blonder, Tinkham,
  and Klapwijk}}]{Blonder1982}
\bibinfo{author}{\bibfnamefont{G.~E.} \bibnamefont{Blonder}},
  \bibinfo{author}{\bibfnamefont{M.}~\bibnamefont{Tinkham}}, \bibnamefont{and}
  \bibinfo{author}{\bibfnamefont{T.~M.} \bibnamefont{Klapwijk}},
  \bibinfo{journal}{Phys. Rev. B} \textbf{\bibinfo{volume}{25}},
  \bibinfo{pages}{4515} (\bibinfo{year}{1982}).

\bibitem[{\citenamefont{Blanter and B{\"{u}}ttiker}(2000)}]{Blanter2000}
\bibinfo{author}{\bibfnamefont{Y.}~\bibnamefont{Blanter}} \bibnamefont{and}
  \bibinfo{author}{\bibfnamefont{M.}~\bibnamefont{B{\"{u}}ttiker}},
  \bibinfo{journal}{Phys. Rep.} \textbf{\bibinfo{volume}{336}},
  \bibinfo{pages}{1} (\bibinfo{year}{2000}).

\bibitem[{\citenamefont{Anantram and Datta}(1996)}]{Anantram1996}
\bibinfo{author}{\bibfnamefont{M.~P.} \bibnamefont{Anantram}} \bibnamefont{and}
  \bibinfo{author}{\bibfnamefont{S.}~\bibnamefont{Datta}},
  \bibinfo{journal}{Phys. Rev. B} \textbf{\bibinfo{volume}{53}},
  \bibinfo{pages}{16390} (\bibinfo{year}{1996}).

\bibitem[{\citenamefont{Coleman and Schofield}(1995)}]{Coleman1995}
\bibinfo{author}{\bibfnamefont{P.}~\bibnamefont{Coleman}} \bibnamefont{and}
  \bibinfo{author}{\bibfnamefont{A.~J.} \bibnamefont{Schofield}},
  \bibinfo{journal}{Phys. Rev. Lett.} \textbf{\bibinfo{volume}{75}},
  \bibinfo{pages}{2184} (\bibinfo{year}{1995}).
%, \eprint{9504084}


\bibitem[{\citenamefont{Coleman et~al.}(1995)\citenamefont{Coleman, Ioffe, and
  Tsvelik}}]{Coleman1995a}
\bibinfo{author}{\bibfnamefont{P.}~\bibnamefont{Coleman}},
  \bibinfo{author}{\bibfnamefont{L.~B.} \bibnamefont{Ioffe}}, \bibnamefont{and}
  \bibinfo{author}{\bibfnamefont{a.~M.} \bibnamefont{Tsvelik}},
  \bibinfo{journal}{Phys. Rev. B} \textbf{\bibinfo{volume}{52}},
  \bibinfo{pages}{6611} (\bibinfo{year}{1995}).

\bibitem[{\citenamefont{Ng and Lee}(1988)}]{Ng1988}
\bibinfo{author}{\bibfnamefont{T.~K.} \bibnamefont{Ng}} \bibnamefont{and}
  \bibinfo{author}{\bibfnamefont{P.~a.} \bibnamefont{Lee}},
  \bibinfo{journal}{Phys. Rev. Lett.} \textbf{\bibinfo{volume}{61}},
  \bibinfo{pages}{1768} (\bibinfo{year}{1988}).

\bibitem[{\citenamefont{Ringel et~al.}(2012)\citenamefont{Ringel, Kraus, and
  Stern}}]{Ringel2012}
\bibinfo{author}{\bibfnamefont{Z.}~\bibnamefont{Ringel}},
  \bibinfo{author}{\bibfnamefont{Y.~E.} \bibnamefont{Kraus}}, \bibnamefont{and}
  \bibinfo{author}{\bibfnamefont{A.}~\bibnamefont{Stern}},
  \bibinfo{journal}{Phys. Rev. B} \textbf{\bibinfo{volume}{86}},
  \bibinfo{pages}{045102} (\bibinfo{year}{2012}).
%, \eprint{1105.4351}


\end{thebibliography}

\appendix
\begin{widetext}
\section{Derivation of the effective SD model for $ \nu=2$}\label{appendice}
We perform a Schrieffer Wolff transformation on the Hamiltonian $H=H_0+H_T$ [Eqs.~\eqref{H:majo} and \eqref{HT}] taking into account virtual transitions to excited states. 
For this purpose we define the following projection operators on the ground-state manifold and on the excited states, respectively:
\be
{\cal P}_g &=& |{\rm gs}_{ +-}\rangle \langle {\rm gs}_{+-}|+ |{\rm gs}_{ -+}\rangle \langle {\rm gs}_{-+}|,\nonumber\\
{\cal P}_e &=& 1 - {\cal P} = \sum_{n, s,s'} |n_{s,s'}\rangle \langle n_{s,s'}|, \nonumber
\ee
where the sum is over excited states $|n_{s,s'}\rangle$, for which $H_0 |n_{s,s'}\rangle =E_n |n_{s,s'}\rangle$, $P_o  |n_{s,s'}\rangle=s   |n_{s,s'}\rangle$, and $ P_e |n_{s,s'}\rangle=s' |n_{s,s'}\rangle$.  
Noting  that tunnelling events from the right lead change the parity on the even subchain while tunnelling events on the left lead change the parity of the odd subchain, the matrix elements of the tunnelling Hamiltonian between the ground-state manifold and excited states are:
\be
H_{ge} &=& {\cal P}_g H{\cal P}_e = {\cal P}_g H_T{\cal P}_e = \sum_k\sum_n \left\{ |{\rm gs}_{ +-}\rangle \langle n_{++}| \left(t_R c_{R,k}^\dag\tilde{\beta}_n+t_R^* c_{R,k} \tilde{\alpha}_n^*
\right) \right. \nonumber\\
& &+|{\rm gs}_{ +-}\rangle \langle n_{--}| \left(t_L c_{L,k}^\dag\beta_n-t_L^* c_{L,k} \alpha_n^*\right)\nonumber\\
& & +|{\rm gs}_{ -+}\rangle \langle n_{++}| \left( t_L c_{L,k}^\dag \alpha_n-t_L^* c_{L,k} \beta_n^* \right) \nonumber\\
& & \left. +|{\rm gs}_{ -+}\rangle \langle n_{--}| \left(t_R c_{R,k}^\dag \tilde{\alpha}_n+t_R^* c_{R,k} \tilde{\beta}_n^*
\right)\right\},\nonumber
\ee
where we have used the relations which follow from charge conjugation symmetry $ T_L$:
\be
\alpha_n &=&\langle {\rm gs}_{ -+}|  d_1|n_{++}\rangle=  \left(\langle {\rm gs}_{ +-}|  d_1^\dag|n_{--}\rangle\right)^*,\nonumber\\
\beta_n &=&\langle {\rm gs}_{ +-}|  d_1|n_{--}\rangle= \left(\langle {\rm gs}_{-+}|  d_1^\dag|n_{++}\rangle\right)^*,\nonumber\\
\tilde{\alpha}_n &=&\langle {\rm gs}_{ -+}|  d_N|n_{--}\rangle=- \left( \langle {\rm gs}_{ +-}|  d_N^\dag|n_{++}\rangle\right)^*,\nonumber\\
\tilde{\beta}_n &=&\langle {\rm gs}_{ +-}|  d_N|n_{++}\rangle=-\left(  \langle {\rm gs}_{-+}|  d_N^\dag|n_{--}\rangle\right)^*.\nonumber
\ee
The resulting effective model  is given by $H_{\rm eff} = H_{ge}(E_g-H_{ee})^{-1}H_{eg} $:
\be
H_{\rm eff}&= & |{\rm gs}_{ +-}\rangle \langle  {\rm gs}_{ +-}| \sum_{k,k'}\sum_n \Bigg\{ |t_R|^2\frac{|\tilde{\beta}_n|^2-|\tilde{\alpha}_n|^2}{E_g-E_n}c_{R,k}^\dag c_{R,k'}+|t_L|^2\frac{|\beta_n|^2-|\alpha_n|^2}{E_g-E_n}c_{L,k}^\dag c_{L,k'} + \left(|t_R|^2\frac{|\tilde{\alpha}_n|^2}{E_g-E_n} + |t_L|^2\frac{|\alpha_n|^2}{E_g-E_n}\right)\delta_{k,k'} \Bigg\}\nonumber \\
&+& |{\rm gs}_{ -+}\rangle \langle  {\rm gs}_{ -+}| \sum_{k,k'}\sum_n \left\{ |t_R|^2\frac{|\tilde{\alpha}_n|^2-|\tilde{\beta}_n|^2}{E_g-E_n}c_{R,k}^\dag c_{R,k'}+|t_L|^2\frac{|\alpha_n|^2-|\beta_n|^2}{E_g-E_n}c_{L,k}^\dag c_{L,k'} + \left(|t_R|^2\frac{|\tilde{\beta_n}|^2}{E_g-E_n} + |t_L|^2\frac{|\beta_n|^2}{E_g-E_n} \right)\delta_{k,k'}\right\} \nonumber\\
&+&|{\rm gs}_{ +-}\rangle \langle  {\rm gs}_{ -+}| \sum_{k,k'}\sum_n \Bigg( t_Rt_L^*\frac{2\tilde{\beta}_n\alpha_n^*}{E_g-E_n}c_{R,k}^\dag c_{L,k'}+t_Rt_L\frac{2\tilde{\beta}_n\beta_n}{E_g-E_n}c_{L,k}^\dag c_{R,k'}^\dag+t_R^*t_L^*\frac{2\tilde{\alpha}_n^*\alpha_n^*}{E_g-E_n}c_{R,k}c_{L,k'}
+t_R^*t_L\frac{2\tilde{\alpha}_n^*\beta_n}{E_g-E_n}c_{L,k}^\dag c_{R,k'}\Bigg)\nonumber\\
&+&{\rm h.c.}\nonumber\\
&=& \tau_z  \sum_{k,k'}\sum_n\left\{ |t_R|^2\frac{|\tilde{\beta}_n|^2-|\tilde{\alpha}_n|^2}{E_g-E_n}c_{R,k}^\dag c_{R,k'}+|t_L|^2\frac{|\beta_n|^2-|\alpha_n|^2}{E_g-E_n} c_{L,k}^\dag c_{L,k'}
-\frac{\delta_{k,k'}}{2}\left(|t_R|^2\frac{|\tilde{\beta}_n|^2-|\tilde{\alpha}_n|^2}{E_g-E_n}+|t_L|^2\frac{|\beta_n|^2-|\alpha_n|^2}{E_g-E_n}\right)\right\} \nonumber\\
&+&\Bigg\{\tau_+\sum_{k,k'}\sum_n \Bigg( t_Rt_L^*\frac{2\tilde{\beta}_n\alpha_n^*}{E_g-E_n}c_{R,k}^\dag c_{L,k'}+t_Rt_L\frac{2\tilde{\beta}_n\beta_n}{E_g-E_n}c_{L,k}^\dag c_{R,k'}^\dag+t_R^*t_L^*\frac{2\tilde{\alpha}_n^*\alpha_n^*}{E_g-E_n}c_{R,k}c_{L,k'}
+t_R^*t_L\frac{2\tilde{\alpha}_n^*\beta_n}{E_g-E_n}c_{L,k}^\dag c_{R,k'}\Bigg)+{\rm h.c.}\Bigg\},\nonumber
\ee
where the last equation is up to an additive constant and we have defined $\tau_z = |{\rm gs}_{ +-}\rangle \langle  {\rm gs}_{ +-}| - |{\rm gs}_{ -+}\rangle \langle  {\rm gs}_{ -+}| $ and $\tau_+=|{\rm gs}_{ +-}\rangle \langle  {\rm gs}_{ -+}|$.
Introducing the coefficients:
\be
J_z &=&  \sum_n\left\{|t_R|^2  \frac{|\tilde{\beta}_n|^2-|\tilde{\alpha}_n|^2}{E_g-E_n}-|t_L|^2  \frac{|\beta_n|^2-|\alpha_n|^2}{E_g-E_n}\right\},\nonumber\\
J_- &=& 2t_Rt_L^*\sum_n\frac{\tilde{\beta}_n\alpha_n^*}{E_g-E_n}\nonumber,\\
J_+ &=&2t_R^*t_L\sum_n \frac{\tilde{\alpha}_n^*\beta_n}{E_g-E_n},\nonumber
\ee
and
\be
\Delta_z&=& \sum_n\left\{|t_R|^2  \frac{|\tilde{\beta}_n|^2-|\tilde{\alpha}_n|^2}{E_g-E_n}+|t_L|^2  \frac{|\beta_n|^2-|\alpha_n|^2}{E_g-E_n}\right\},\nonumber\\
\Delta_+ &=&2t_Rt_L\sum_n\frac{\tilde{\beta}_n\beta_n}{E_g-E_n},\nonumber\\
\Delta_-&=&2t_R^*t_L^*\sum_n\frac{\tilde{\alpha}_n^*\alpha_n^*}{E_g-E_n}, \nonumber
\ee
we arrive at the expression given in Eq.~\eqref{H_SD} of the main text. 
\end{widetext}

\end{document}